\documentclass{article}

\usepackage{spconf}
\usepackage{booktabs}
\usepackage{amsmath,amssymb,amsfonts}
\usepackage{graphicx}
\usepackage{cite}
\usepackage{amsmath,amssymb,amsfonts}
\usepackage{bm, enumitem}
\usepackage{xcolor}
\usepackage{acronym}
\usepackage{makecell}
\usepackage{xurl}
\usepackage[bookmarks,colorlinks=true]{hyperref}
\usepackage{algorithmic} 
\usepackage[ruled,linesnumbered,vlined]{algorithm2e}

 \usepackage[all=normal,paragraphs=tight,floats=normal,mathspacing=normal,wordspacing=tight,charwidths=tight,mathdisplays=normal,leading=normal]{savetrees}
\acrodef{pnp}[PnP]{plug-and-play}
\acrodef{cc}[CC]{Compton camera}
\acrodef{hls}[HLS]{homeland security}
\acrodef{em}[EM]{expectation maximization}
\acrodef{mlem}[MLEM]{maximum likelihood expectation maximization}
\acrodef{map}[MAP]{maximum a posteriori}
\acrodef{fwhm}[FWHM]{full width at half maximum}
\acrodef{ptvr}[PTVR]{peak-to-valley ratio}
\acrodef{gamos}[GAMOS]{GEANT4-based Architecture for Medicine-Oriented Simulations}
\acrodef{nn}[NN]{neural network}
\acrodef{cnn}[CNN]{convolutional neural network}
\acrodef{nmse}[NMSE]{normalized mean-square error}
\acrodef{ssim}[SSIM]{structural similarity index measure}
\acrodef{psnr}[PSNR]{peak signal-to-noise ratio}
\acrodef{rl}[RL]{Richardson-Lucy}

\newcommand{\vect}[1]{\bm{#1}}

\title{Deep-Unfolded Penalized MLEM for Rapid Poisson Image Reconstruction}

\name{
Zohar Davidov, 
Alon Osovizky, 
Max Ghelman, and 
Nir Shlezinger
\thanks{
 Z. Davidov and N. Shlezinger are with the ECE School, Ben-Gurion University of the Negev, Israel (e-mail: zohardav@post.bgu.ac.il; nirshl@bgu.ac.il). A. Osovizky and M. Ghelman are with the Nuclear Research Center Negev, Beer-Sheva, Israel (e-mail: alon@rotemi.co.il; maxghelman@gmail.com)
}}
\address{\vspace{-25mm}}

\begin{document}
\ninept
\maketitle

\begin{abstract} 
\Ac{mlem} is a common approach for Poisson image reconstruction, but accurate recovery requires many iterations and is sensitive to mismatch in the assumed acquisition model. We propose a deep-unfolded penalized \ac{mlem} framework that maps a prescribed small number of iterations into trainable layers while retaining the analytical forward/backward projections and the multiplicative structure of \ac{mlem}. The unfolded architecture learns layer-dependent regularization and Poisson-model parameters, together with a data-dependent correction of the sensitivity normalization to mitigate model mismatch. We numerically show that our method improves reconstruction resolution at a fixed iteration budget and attains reconstruction quality comparable to long \ac{mlem} runs and direct data-driven models.

\end{abstract}

\begin{keywords}
deep unfolding, Poisson image reconstruction, maximum likelihood expectation maximization, Compton camera
\end{keywords}

\acresetall

\vspace{-0.1cm}
\section{Introduction}  
\vspace{-0.1cm}
Photon-counting imaging systems arise in a broad range of applications, including emission tomography, fluorescence microscopy,   and gamma-ray imaging. In such systems, the acquired measurements are naturally described by Poisson statistics, leading to inverse problems in which an image is recovered from noisy photon counts through a known or approximately known forward operator~\cite{caucci2016radiance}.
A classical approach to such problems is \ac{mlem}, which maximizes the Poisson likelihood through a sequence of iterations~\cite{shepp1982maximum}. Despite its principled statistical formulation, \ac{mlem} may require many iterations to achieve accurate reconstructions, while under model-mismatched settings, continued iterations can even amplify measurement noise \cite{barrett1994noiseI,wilson1994noiseII}. Penalized-likelihood variants mitigate this effect by incorporating prior structure into the reconstruction objective~\cite{green1990bayesian, fessler1995penalized}, but introduce regularization parameters that must be selected and typically remain fixed throughout the iterative procedure.

Various model-based approaches have been developed to reduce the computational burden of iterative reconstruction under Poisson observations. A common direction accelerates the underlying likelihood optimization by modifying how the measurements or unknown variables are processed. Examples include the space-alternating generalized \ac{em} algorithm~\cite{fessler1994space}, ordered-subsets \ac{em}~\cite{hudson1994accelerated}, and row-action maximum-likelihood methods~\cite{browne1996row}, as well as subsequent relaxed and block-sequential schemes designed to retain fast initial convergence while improving convergence guarantees~\cite{ahn2003globally}. Recent acceleration strategies include simultaneous coordinate-ascent updates~\cite{le2026bayesian} and iteration-dependent preconditioning~\cite{guo2022fast}. A complementary line of work replaces the classical \ac{em} update by optimization procedures tailored to regularized Poisson likelihoods, using, e.g., sparse and total-variation  priors~\cite{harmany2012spiral} or higher-order spatial  regularization~\cite{lefkimmiatis2013poisson}.
Complexity can also be reduced by exploiting structure in the acquisition model, such as measurement aggregation~\cite{davidov2025grouping}.
Despite these advances, the reconstruction updates and their iterative behavior are largely prescribed analytically relying on known forward operator and numerous iterations.

Model-based deep learning provides a means to preserve the structure of the physical model while adapting selected algorithmic components from
data~\cite{shlezinger2023model}. One prominent realization is \ac{pnp} reconstruction, which incorporates learned priors through denoising operators embedded within iterative optimization algorithms~\cite{kamilov2023plug}. \ac{pnp} methods have  been adapted to Poisson inverse problems in~\cite{rond2016poisson, hurault2023convergent,klatzer2026efficient}. A complementary model-based learning strategy is deep unfolding, which maps a prescribed number of iterations into trainable layers and thereby enables direct optimization of the finite-iteration reconstruction procedure~\cite{Shlezinger2025DeepUnfolding}. This principle has been applied to imaging inverse problems~\cite{liu2021stochastic}, including Poisson deconvolution, although the latter uses a learned weighted least-squares surrogate in place of the Poisson data-fidelity term~\cite{singh2025learning}. %Related \ac{em}-based approaches have incorporated learned regularizers into positron-emission-tomography reconstruction by unfolding maximum-a-posteriori \ac{em} updates~\cite{mehranian2021model,corda2021memory}. These works demonstrate the benefits of combining learned image priors and trainable optimization with Poisson data fidelity, and highlight the opportunity to go beyond learning only the image prior or generic optimization parameters, adapting the finite-iteration dynamics of structured Poisson reconstruction algorithms themselves.
 
In this work, we propose a deep-unfolded penalized \ac{mlem} framework for Poisson computational imaging. Our architecture maps a prescribed small number of \ac{mlem} iterations into trainable layers while retaining the analytical Poisson forward and backward projections and the nonnegative multiplicative structure of the classical update. Unlike conventional penalized \ac{mlem}, whose reconstruction rule is fixed across iterations, our formulation learns an iteration-dependent regularization parameter, together with a data-dependent correction of the nominal sensitivity normalization that enables the unfolded iterations to compensate for inaccuracies in the assumed imaging model. 
We evaluate the proposed method on \ac{rl} deconvolution~\cite{lucy1974iterative,richardson1972bayesian} and \ac{cc} imaging~\cite{kim2024review}, comparing it with direct purely-data-driven models adapted from DenseDeconNet~\cite{Guo2020DenseDeconNet} and ComptonNet~\cite{Sato2024ComptonNet}, respectively. In both experimental setups, our unfolded architecture improves reconstruction quality over these direct models and achieves this with a fixed number of layers, compared with substantially longer classical \ac{mlem} and penalized-\ac{mlem} algorithms.

% \smallskip
% \noindent
 %{\bf Notations:} We use  boldface lowercase for vectors, e.g., $\mathbf{x}$, and boldface uppercase letters, e.g., $\mathbf{M}$, for matrices, with $[\mathbf{x}]_n$ being the $n$th element of $\mathbf{x}$. We use  $\mathbb{R}$ for the set of real numbers. The operations $(\cdot)^{\mathsf T}$ and  $\| \cdot \|$  are used for transpose and  $\ell_2$ norm,  respectively. % while $\operatorname{Poisson}(\cdot)$ is the Poisson distribution. 

%---------------------------------------
%	Poisson Imaging and MLEM
%---------------------------------------
\vspace{-0.1cm}
\section{Poisson Imaging and MLEM}
\label{sec:preliminaries}
\vspace{-0.1cm}

{\bf Signal Model:}
We consider a photon-counting  system where an image $\mathbf{x}\in\mathbb{R}_{+}^{N}$ is observed through a nonnegative  forward operator $\mathbf{A}\in\mathbb{R}_{+}^{M\times N}$. The entries of the  observed  $\mathbf{y}\in\mathbb{N}_{+}^{M}$ obey a Poisson model
\begin{equation}
    [\mathbf{y}]_m \sim \operatorname{Poisson}\left( [\mathbf{A}\mathbf{x}]_m+[\mathbf{b}]_m    \right),
    \qquad m=1,\ldots,M,
    \label{eq:poisson_model}
\end{equation}
where $\mathbf{b}\in\mathbb{R}_{+}^{M}$ represents a known background contribution. The corresponding negative log-likelihood (up to terms independent of $\mathbf{x}$) is
\begin{equation}
    -\log \Pr(\mathbf{y}|\mathbf{x})  \propto \mathbf{1}^{\mathsf T} (\mathbf{A}\mathbf{x}+\mathbf{b}) - \mathbf{y}^{\mathsf T}     \log\left(\mathbf{A}\mathbf{x}+\mathbf{b}\right),
    \label{eq:poisson_nll}
\end{equation}
where the logarithm is applied element-wise.

\smallskip
\noindent
{\bf Problem Formulation:}
Our goal is to recover $\mathbf{x}$ from $\mathbf{y}$ using the physical forward operator $\mathbf{A}$, while focusing on the following challenges:
\begin{enumerate}[label={C\arabic*}]
    \item \label{itm:slow_mlem}
    Recovery must be carried out within a limited latency.
    \item \label{itm:model_mismatch}
    The available $\mathbf{A}$ may only approximate the acquisition model.
\end{enumerate}
To cope with \ref{itm:slow_mlem}-\ref{itm:model_mismatch}, we assume access to a dataset of measurement and ground-truth image pairs, denoted $\mathcal{D}=
\big\{(\vect{y}_i,\vect{x}^\star_i)\big\}_{i=1}^{|\mathcal{D}|}$.

\smallskip
\noindent
{\bf MLEM Reconstruction:}
A classical approach for minimizing~\eqref{eq:poisson_nll} under the nonnegativity constraint is \ac{mlem}~\cite{shepp1982maximum}. Defining the sensitivity vector  $\mathbf{s}=\mathbf{A}^{\mathsf T}\mathbf{1}$,
and initializing with $\mathbf{x}^{(0)}>\mathbf{0}$, its $k$th update is
\begin{equation}
    \mathbf{x}^{(k+1)} = \mathbf{x}^{(k)} \odot \frac{ \mathbf{A}^{\mathsf T}\mathbf{p}^{(k)}  } {\mathbf{s}}, 
    \qquad 
    \mathbf{p}^{(k)} =   \dfrac{\mathbf{y}} {\mathbf{A}\mathbf{x}^{(k)}+\mathbf{b}} ,
    \label{eq:mlem}
\end{equation}
where $\odot$ and all divisions denote element-wise operations. Each iteration thus requires one forward projection through $\mathbf{A}$ and one backward projection through $\mathbf{A}^{\mathsf T}$. Consequently,~\ref{itm:slow_mlem} becomes particularly pronounced when a large number of \ac{mlem} iterations is required.

To improve convergence in spatially extended  settings, \ac{mlem} can be augmented with a spatial regularizer~\cite{green1990bayesian,fessler1995penalized}, e.g., a quadratic penalty
\begin{equation}
    \mathcal{R}(\mathbf{x})  = \frac{1}{4} \sum_{n=1}^{N} \sum_{q\in\mathcal{N}(n)} ([\mathbf{x}]_n-[\mathbf{x}]_q)^2,
    \label{eq:penalty}
\end{equation}
where $\mathcal{N}(n)$ denotes the neighborhood of pixel $n$. Its gradient admits the decomposition 
   $ \nabla\mathcal{R}(\mathbf{x})  =  \mathbf{u}(\mathbf{x})-\mathbf{v}(\mathbf{x})$, 
with $[\mathbf{u}]_n(\mathbf{x})=d_n [\mathbf{x}]_n$ and $[\mathbf{v}]_n(\mathbf{x})=\sum_{q\in\mathcal{N}(n)} [\mathbf{x}]_q$, where $d_n=|\mathcal{N}(n)|$.
The resulting penalized \ac{mlem} update is
\begin{equation}
    \mathbf{x}^{(k+1)} = \mathbf{x}^{(k)} \odot \frac{  \mathbf{A}^{\mathsf T} \mathbf{p}^{(k)} +  \gamma\mathbf{v}(\mathbf{x}^{(k)})}{\mathbf{s}  + \gamma\mathbf{u}(\mathbf{x}^{(k)}) },
    \label{eq:penalized_mlem}
\end{equation}
where $\gamma>0$ controls the trade-off between data fidelity and spatial smoothness. Although regularization suppresses noise amplification, conventional penalized \ac{mlem} uses a fixed coefficient $\gamma$ and an identical analytical update at every iteration, limiting its ability to adapt across iterations.  Moreover, both~\eqref{eq:mlem} and~\eqref{eq:penalized_mlem} rely explicitly on the assumed operator $\mathbf{A}$ and the sensitivity $\mathbf{s}$ derived from it, making them susceptible to~\ref{itm:model_mismatch}.

\vspace{-0.1cm}
\section{Deep-Unfolded MLEM}
\label{sec:method}
\vspace{-0.1cm}
The limited-latency requirement in~\ref{itm:slow_mlem} motivates replacing the potentially long \ac{mlem} process with a prescribed small number of  steps. Deep unfolding provides a natural means to do so while retaining the Poisson imaging model~\cite{Shlezinger2025DeepUnfolding}. Moreover, the availability of $\mathcal{D}$ allows the finite reconstruction trajectory to be optimized for the task at hand and to compensate for inaccuracies in the acquisition model~(\ref{itm:model_mismatch}). Accordingly, we unfold penalized \ac{mlem} into a fixed-depth discriminative machine learning architecture~\cite{shlezinger2022discriminative} (Subsection~\ref{ssec:architecture}), where selected model and regularization parameters vary across layers, and a learned correction adapts the sensitivity normalization. We describe its training in Subsection~\ref{ssec:training} and discuss its properties and complexity in Subsection~\ref{ssec:discussion}.

\begin{figure*}[t]
    \centering
    \includegraphics[width=\textwidth]{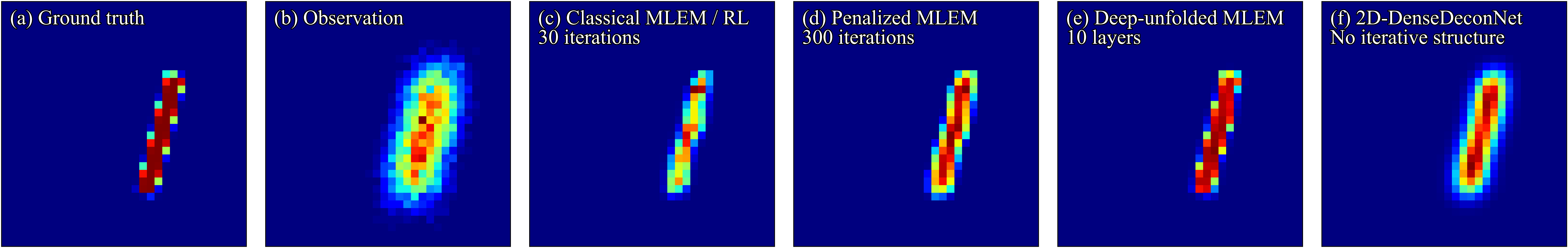}
    \caption{RL deconvolution: (a) ground truth; (b) blurred Poisson observation; (c) MLEM/RL after 30 iterations; (d) penalized MLEM after 300 iterations; (e) unfolded MLEM after 10 layers; (f) direct 2D-DenseDeconNet.}
    \label{fig:rl-thin-rectangle}
\end{figure*}

\begin{figure}[t]
    \centering
    \includegraphics[width=1.0\columnwidth]{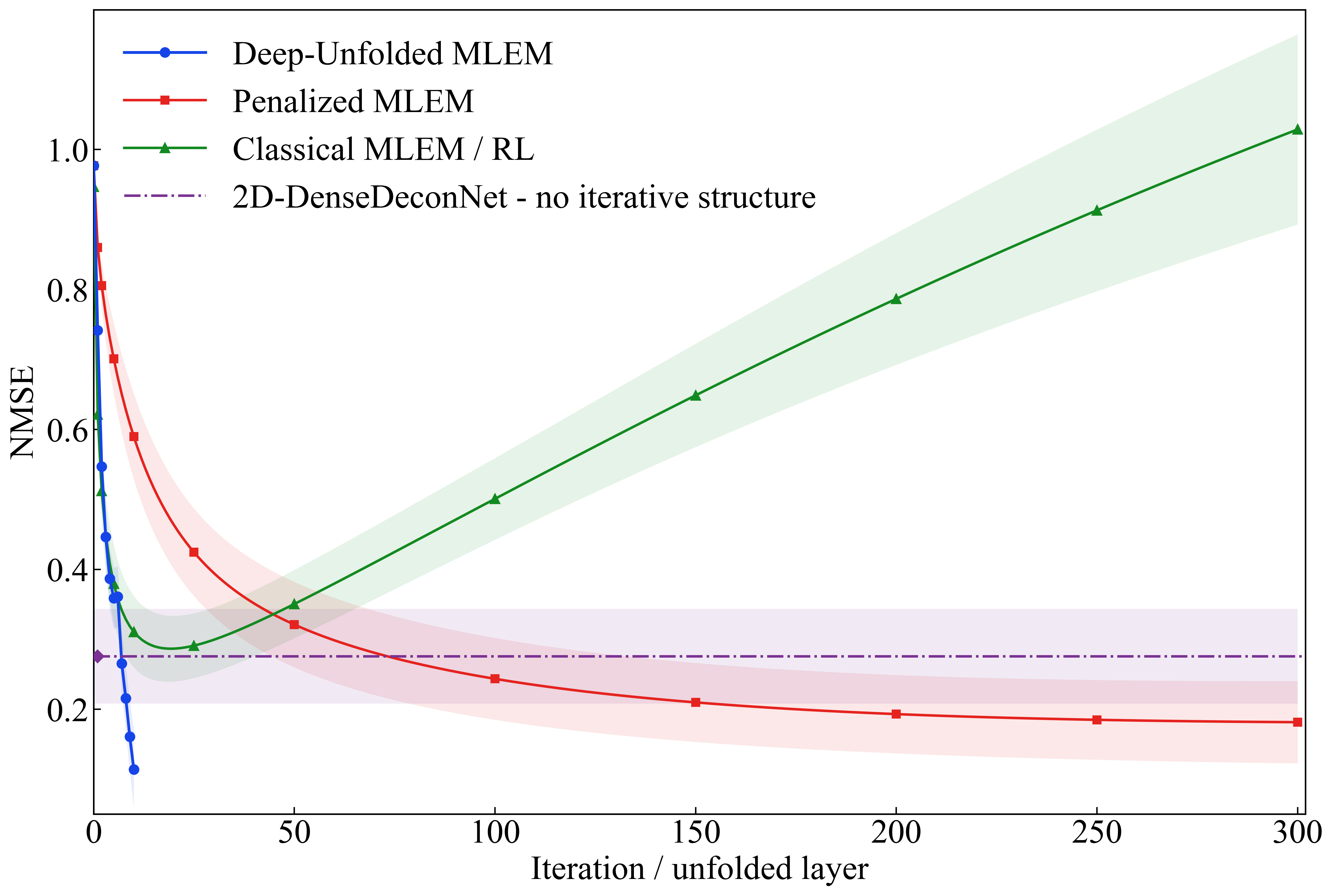}
    \caption{RL NMSE versus iteration (mean $\pm1$ standard deviation). The horizontal 2D-DenseDeconNet line denotes single-pass performance. %Classical methods use 300 iterations, while the unfolded method uses 10 layers.
    }
    \label{fig:rl-convergence}
\end{figure}

\begin{figure*}[t]
    \centering
    \includegraphics[width=\textwidth]{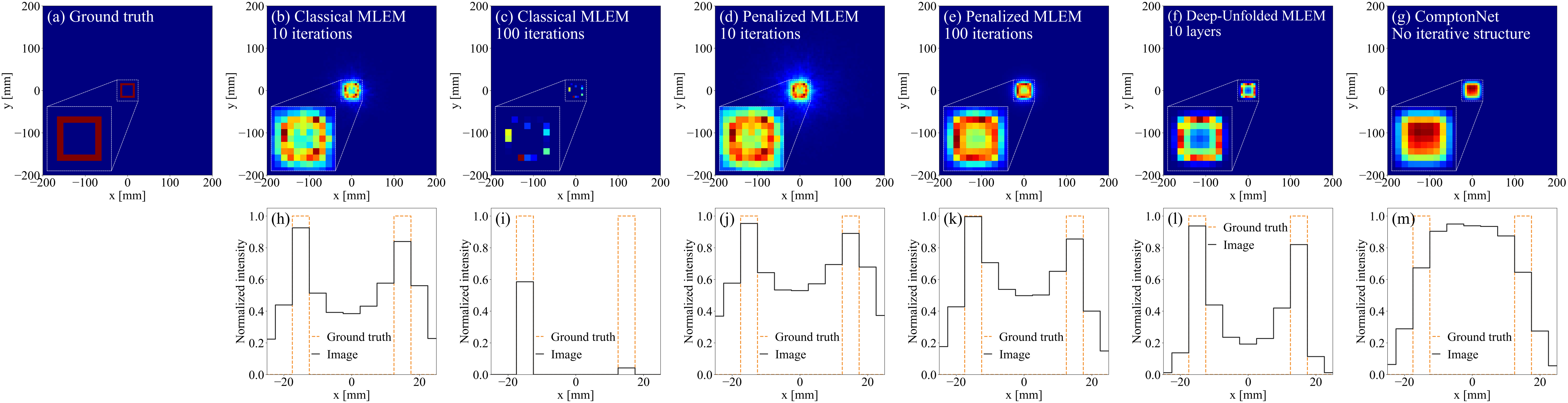}
    \caption{Compton-camera rectangle: (a) ground truth; (b)-(c) MLEM and (d)-(e) penalized MLEM at 10 and 100 iterations; (f) unfolded MLEM, 10 layers; (g) direct ComptonNet. Panels (h)-(m) show their $y=0$ profiles over the central zoom region, with ground truth dashed orange and reconstructions solid black.}
    \label{fig:cc-rectangle-reconstructions}
\end{figure*}

\begin{figure}[t]
    \centering
    \includegraphics[width=1.0\columnwidth]{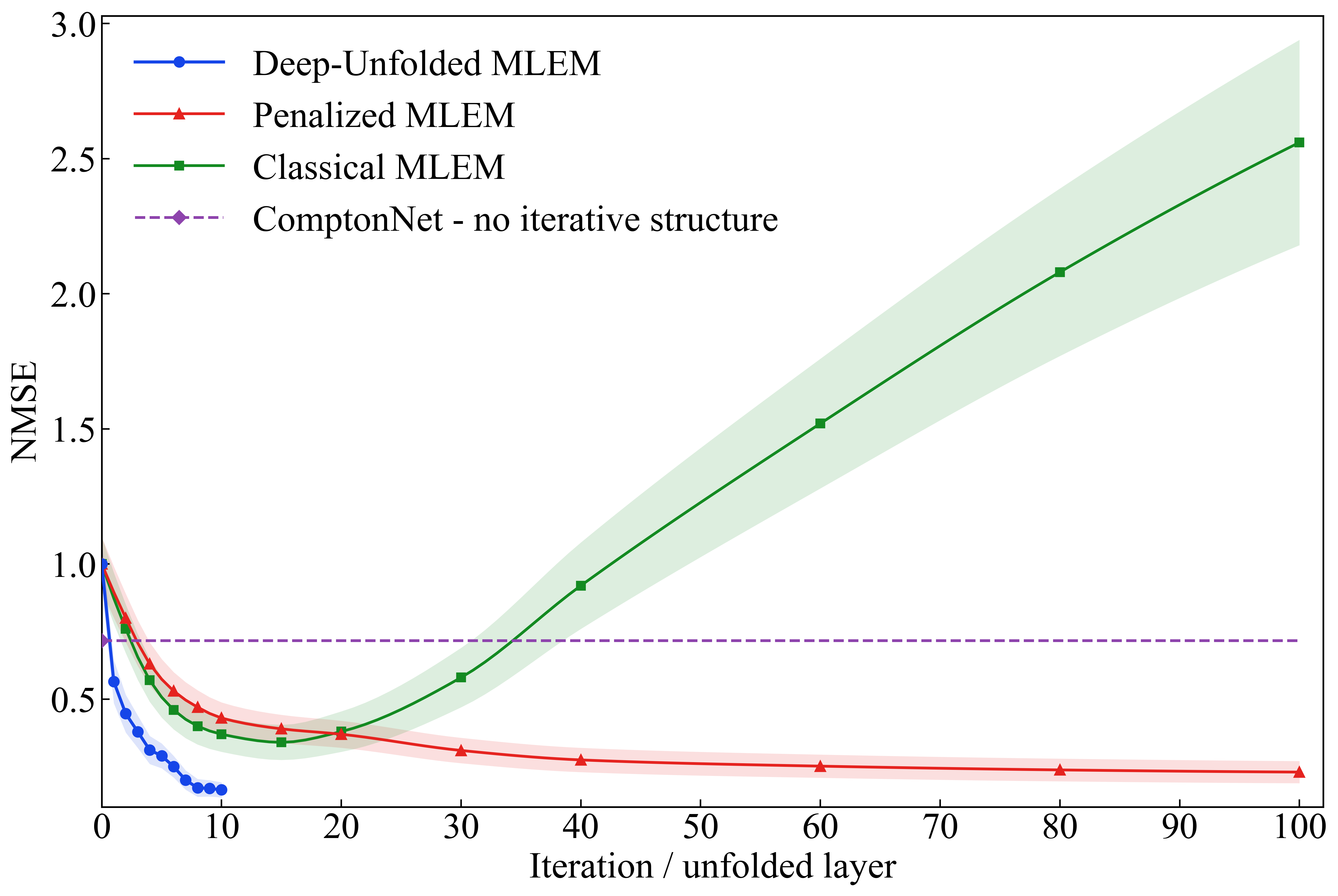}
    \caption{CC NMSE versus iteration (mean $\pm1$ standard deviation). The horizontal ComptonNet line denotes single-pass performance.}
    \label{fig:cc-nmse-convergence}
\end{figure}

%----------------------------------------------------------------
\vspace{-0.1cm}
\subsection{Architecture}
\label{ssec:architecture}
\vspace{-0.1cm}

% \noindent
% {\bf Unfolded MLEM:}
We unfold $K$ iterations of~\eqref{eq:penalized_mlem}, representing each iteration as a trainable layer. The $k$th layer receives $\mathbf{x}^{(k)}$ and retains the analytical forward and backward operators $\mathbf{A}$ and $\mathbf{A}^{\mathsf T}$, while allowing the quantities governing the reconstruction update to vary across layers.

In particular, let $\alpha_k>0$ denote a layer-dependent scaling of the predicted photon intensity and $\beta_k>0$ scale the nominal background $\mathbf{b}$. %, and  $\gamma_k>0$ be a layer-dependent regularization coefficient. 
The predicted  likelihood backprojection  at iteration $k$ is set to
\begin{align}
    %\mathbf{r}^{(k)} &=
   % \alpha_k\mathbf{A}\mathbf{x}^{(k)} +\beta_k\mathbf{b},
   % \qquad 
    \tilde{\mathbf{p}}^{(k)} =
    \mathbf{A}^{\mathsf T}\left(\frac{\mathbf{y}}{\alpha_k\mathbf{A}\mathbf{x}^{(k)} +\beta_k\mathbf{b}}\right).
    \label{eq:Ptilde}
\end{align}
Thus, $\alpha_k$ and $\beta_k$ allow the effective Poisson model employed by the finite-depth reconstruction to adapt across layers.  %while $\gamma_k$ controls the corresponding balance between data fidelity and spatial regularization.

% \smallskip
% \noindent
% {\bf Sensitivity Correction:}
The sensitivity $\mathbf{s}=\mathbf{A}^{\mathsf T}\mathbf{1}$ appearing in \eqref{eq:mlem}-\eqref{eq:penalized_mlem} is determined by the assumed acquisition operator. In particular, letting $\mathbf{A}_{\rm true}=\mathbf{A}+\Delta\mathbf{A}$ denote the true operator gives
$    \mathbf{s}_{\rm true}= \mathbf{A}_{\rm true}^{\mathsf T}\mathbf{1}
    = \mathbf{s} +  \Delta\mathbf{A}^{\mathsf T}\mathbf{1}$.
Consequently, model mismatch induces a structured error in the normalization of each \ac{mlem} update. Rather than estimating $\Delta\mathbf{A}$ explicitly, which may be high-dimensional, we learn a low-complexity correction directly in the reconstruction domain.
To this end, the current estimate, likelihood backprojection, and regularizer are processed by a small \ac{cnn} with trainable parameters $\boldsymbol{\psi}_k$  into
\begin{equation}
    \mathbf{c}^{(k)} =
    f_{\boldsymbol{\psi}_k}\left(\mathbf{x}^{(k)}, \tilde{\mathbf{p}}^{(k)},     \mathbf{u}(\mathbf{x}^{(k)}), \mathbf{v}(\mathbf{x}^{(k)}) \right).
    \label{eq:sensitivity_correction}
\end{equation}
%where $\boldsymbol{\psi}_k$ denotes the trainable parameters of the $k$th layer.
 
The resulting unfolded update follows the penalized \ac{mlem} of \eqref{eq:penalized_mlem}, with the scaled backprojection of~\eqref{eq:Ptilde}, the sensitivity correction of~\eqref{eq:sensitivity_correction}, and while allowing iteration-varying regularization parameterized by $\gamma_k$. At iteration $ k=0,\ldots,K-1$, the update rule is
\begin{equation}
\mathbf{x}^{(k+1)} = \mathbf{x}^{(k)} \odot \frac{     \alpha_k\tilde{\mathbf{p}}^{(k)}+ \gamma_k\mathbf{v}(\mathbf{x}^{(k)})}
{\alpha_k\mathbf{s} +\mathbf{c}^{(k)} + \gamma_k\mathbf{u}(\mathbf{x}^{(k)})
}.
\label{eq:unfolded_mlem}
\end{equation}
The parameters $\{\alpha_k,\beta_k,\gamma_k\}$ are constrained to be positive, while the sensitivity correction is constrained such that the denominator in~\eqref{eq:unfolded_mlem} remains positive. After $K$ layers, the reconstruction is set to $\hat{\mathbf{x}}=\mathbf{x}^{(K)}$, as summarized in Algorithm~\ref{alg:unfolded_mlem}. Thus, each unfolded layer retains the multiplicative structure and  projections of penalized \ac{mlem}, while learning how its finite-depth update should adapt to the current recovery state.

\begin{algorithm}
\caption{Deep-Unfolded MLEM Reconstruction}
\label{alg:unfolded_mlem}
\SetKwInOut{Initialization}{Init}
\Initialization{trained $\{\alpha_k,\beta_k,\gamma_k,\boldsymbol{\psi}_k\}_{k=0}^{K-1}$; initial image $\mathbf{x}^{(0)}$
}

\SetKwInOut{Input}{Input}
\Input{measurement $\mathbf{y}$; forward operator $\mathbf{A}$; background $\mathbf{b}$}

Compute the nominal sensitivity $\mathbf{s}\leftarrow\mathbf{A}^{\mathsf T}\mathbf{1}$\;

\For{$k=0,\ldots,K-1$}{
%Compute  predicted intensity $\mathbf{r}^{(k)} \leftarrow \alpha_k\mathbf{A}\mathbf{x}^{(k)} +\beta_k\mathbf{b}$\;

Set likelihood backprojection $\tilde{\mathbf{p}}^{(k)}$ via \eqref{eq:Ptilde}\;

Compute the regularization terms $\mathbf{u}(\mathbf{x}^{(k)})$ and $\mathbf{v}(\mathbf{x}^{(k)})$\;

Compute the sensitivity correction
$\mathbf{c}^{(k)}$ via \eqref{eq:sensitivity_correction}\;

Update $\mathbf{x}^{(k+1)}$ via \eqref{eq:unfolded_mlem}\;
}

\KwRet{Reconstructed image $\hat{\mathbf{x}}=\mathbf{x}^{(K)}$}
\end{algorithm}

%----------------------------------------------------------------
\vspace{-0.1cm}
\subsection{Training}
\label{ssec:training}
\vspace{-0.1cm}
We train the unfolded estimator using the dataset $\mathcal D$. Let $\boldsymbol{\theta}_k=\{\alpha_k,\beta_k,\gamma_k,\boldsymbol{\psi}_k\}$ collect the parameters of layer $k$, and let $\boldsymbol{\theta}=\{\boldsymbol{\theta}_k\}_{k=0}^{K-1}$. Training is performed in two stages, combining progressive initialization with end-to-end optimization.

\smallskip
\noindent
{\bf Progressive Initialization:}
We first train the layers sequentially. For layer $k$, the preceding layers are kept fixed and $\boldsymbol{\theta}_k$ is optimized according to
\begin{equation}
    \mathcal{L}_{\mathcal D}^{(k)}(\boldsymbol{\theta}_k) =  \frac{1}{|\mathcal D|} \sum_{i=1}^{|\mathcal D|} \ell\left(    \mathbf{x}^{(k+1)}_i,\mathbf{x}^\star_i  \right),
    \label{eq:progressive_loss}
\end{equation}
where $\mathbf{x}^{(k+1)}_i$ denotes the reconstruction produced after the newly introduced layer, and $\ell(\cdot,\cdot)$ is a hybrid reconstruction loss defined as
\begin{align} 
&\ell\!\left(
\mathbf{x}^{(k+1)}_i,\mathbf{x}^{\star}_i
\right)
= 
\underbrace{
-\lambda_{\mathrm{shape}}
\sum_{n=1}^{N}
\left[\overline{\mathbf{x}}^{\star}_i\right]_n
\log\!\left(
\left[\overline{\mathbf{x}}^{(k+1)}_i\right]_n
\right)
}_{\text{image-shape agreement}}
\notag \\
&+\!
\underbrace{
\lambda_{\mathrm{loc}}
\left\|
\widehat{\boldsymbol{\mu}}^{(k+1)}_i
\!-\!
\boldsymbol{\mu}^{\star}_i
\right\|_2^2
}_{\text{source localization accuracy}}
 +
\underbrace{
\lambda_{\mathrm{data}}
\left[
\mathbf{1}^{\mathsf T}\mathbf{r}^{(k)}_i
\!-\!
\mathbf{y}_i^{\mathsf T}
\log\!\left(
\mathbf{r}^{(k)}_i
\right)
\right]
}_{\text{Poisson data consistency}} . 
\label{eq:sample_hybrid_loss}
\end{align}
Here, $\lambda_{\mathrm{shape}}\geq0$, $\lambda_{\mathrm{loc}}\geq0$, and $\lambda_{\mathrm{data}}\geq0$ are weighting coefficients controlling the relative contributions of the image-shape, localization, and Poisson data-consistency terms, respectively. The reconstructed and ground-truth centroids are given by $\widehat{\boldsymbol{\mu}}^{(k+1)}_i
=
\sum_{n=1}^{N}
\big[
\overline{\mathbf{x}}^{(k+1)}_i
\big]_n
\boldsymbol{\rho}_n, \ 
\boldsymbol{\mu}^{\star}_i
=
\sum_{n=1}^{N}
\left[
\overline{\mathbf{x}}^{\star}_i
\right]_n
\boldsymbol{\rho}_n ,$ respectively, where $\boldsymbol{\rho}_n$ denotes the physical coordinate of pixel $n$. The corresponding normalized reconstructed and ground-truth images are $ \overline{\mathbf{x}}^{(k+1)}_i
=
\frac{\mathbf{x}^{(k+1)}_i}
{\mathbf{1}^{\mathsf T}\mathbf{x}^{(k+1)}_i},
\
\overline{\mathbf{x}}^{\star}_i
=
\frac{\mathbf{x}^{\star}_i}
{\mathbf{1}^{\mathsf T}\mathbf{x}^{\star}_i},
$
respectively, and $\mathbf{r}^{(k)}_i = \alpha_k\mathbf{A}\mathbf{x}^{(k+1)}_i +\beta_k\mathbf{b}$ is the predicted measurement intensity. This progressive procedure encourages each layer to provide a useful reconstruction refinement before the complete architecture is jointly optimized.

\smallskip
\noindent
{\bf End-to-End Fine-Tuning:}
Following the progressive initialization, all unfolded layers are jointly fine-tuned using the final reconstruction,
\begin{equation}
    \mathcal{L}_{\mathcal D}(\boldsymbol{\theta}) = \frac{1}{|\mathcal D|}    \sum_{i=1}^{|\mathcal D|} \ell\left( \mathbf{x}^{(K)}_i,\mathbf{x}^\star_i    \right).
    \label{eq:e2e_loss}
\end{equation}
Backpropagation through the unfolded iterations allows the individual layers to specialize while jointly optimizing the reconstruction obtained at the prescribed depth $K$. The empirical risks in \eqref{eq:progressive_loss} and \eqref{eq:e2e_loss} are minimized using the standard Adam optimizer.

%----------------------------------------------------------------
\vspace{-0.1cm}
\subsection{Discussion}
\label{ssec:discussion}
\vspace{-0.1cm}

The proposed architecture addresses Challenges~\ref{itm:slow_mlem}-\ref{itm:model_mismatch} while preserving the structure of Poisson likelihood-based reconstruction. Its prescribed depth $K$ directly bounds the number of forward and backward projections required at inference, addressing~\ref{itm:slow_mlem}. The learned layer-dependent parameters adapt the reconstruction trajectory from data, while the sensitivity correction provides a structured mechanism for mitigating discrepancies between the nominal and actual acquisition models, addressing~\ref{itm:model_mismatch}. In particularly, learning augments rather than replaces the imaging model: $\mathbf{A}$ and $\mathbf{A}^{\mathsf T}$ remain explicitly embedded in every layer.

\smallskip
\noindent
{\bf Complexity Analysis:}
%Let $\operatorname{nnz}(\mathbf{A})$ be the number of nonzero entries of the forward operator, and let $C_{\rm CNN}$ denote the complexity of one evaluation of~\eqref{eq:sensitivity_correction}. Each unfolded layer requires one forward and one backward projection, with combined complexity $\mathcal{O}(\operatorname{nnz}(\mathbf{A}))$, together with $\mathcal{O}(N)$ operations for the  quadratic regularizer and $C_{\rm CNN}$ operations for the learned correction. The overall inference complexity is thus
% \begin{equation}
%     \mathcal{O}\left( K\left( \operatorname{nnz}(\mathbf{A})+N+C_{\rm CNN}    \right) \right).
%     \label{eq:complexity}
% \end{equation}
% A conventional penalized-\ac{mlem} reconstruction using $T$ iterations has complexity $\mathcal{O}\left(T(\operatorname{nnz}(\mathbf{A})+N)\right)$. Thus, an unfolded layer incurs the additional cost of the small \ac{cnn}, but the overall computation can be substantially reduced when the learned architecture achieves the desired reconstruction quality with $K\ll T$.
Table~\ref{tab:inference_complexity} summarizes the inference complexity of the considered architectures, where $T$ denotes the number of conventional \ac{mlem} iterations. Each iteration or unfolded layer applies the forward operator $\mathbf{A}$ and its adjoint $\mathbf{A}^{\mathsf T}$, with dominant complexity $\mathcal{O}(MN)$. The CNN cost is $\mathcal{O}\!\left(Nq^2H(C_{\mathrm{in}}+C_{\mathrm{out}})\right)$, where $q$, $H$, $C_{\mathrm{in}}$, and $C_{\mathrm{out}}$ denote its kernel width, hidden width, and input and output channel counts, respectively. Since these are fixed architectural parameters, the CNN complexity scales as $\mathcal{O}(N)$. Therefore, the unfolded architecture can reduce the overall inference cost when $K\ll T$, as also shown in the runtimes reported in Table~\ref{tab:inference_complexity}.

\begin{table}
\centering

\caption{Inference complexity and runtime measured in \ac{cc} experiment.}
\label{tab:inference_complexity}

\renewcommand{\arraystretch}{1.15}
\setlength{\tabcolsep}{7pt}
\resizebox{\columnwidth}{!}{%
\begin{tabular}{lcc}
\toprule
\textbf{Algorithm} & \textbf{Complexity} & \textbf{Runtime} \\
\midrule

 MLEM
& $\mathcal{O}(TMN)$
& {1617.2 s (100 iterations)} \\

Penalized MLEM
& $\mathcal{O}\!\left(T(MN+N)\right)$
& {1673.7 s  (100 iterations)} \\

Deep-unfolded MLEM
& $\mathcal{O}\!\left(K(MN+N)\right)$
& {181.5 s  (10 layers)} \\

ComptonNet
& $\mathcal{O}\!\left(M+N\right)$
& {3.6 s  (no iterative structure)} \\

\bottomrule
\end{tabular}
}
\end{table}

\vspace{-0.1cm}
\section{Experimental Study}
\label{sec:experiments}
\vspace{-0.1cm}

We evaluate the proposed deep-unfolded \ac{mlem} in two Poisson imaging settings with  different forward operators\footnote{Our code is available at \url{https://github.com/Zohardav/deep-unfolded-mlem}}: synthetic \ac{rl} deconvolution and \ac{cc} gamma-ray imaging.
In the former, $\mathbf{A}$ represents convolution with a point-spread function, such that \ac{mlem} reduces to the generalized \ac{rl} iteration~\cite{richardson1972bayesian,lucy1974iterative}; in the latter, $\mathbf{A}$ is determined by the \ac{cc} acquisition geometry and maps candidate source locations to detected events~\cite{wilderman2001improved,
kim2024review}. We compare classical and penalized \ac{mlem}, the proposed method, a 2D-DenseDeconNet adaptation for RL~\cite{Guo2020DenseDeconNet}, and ComptonNet for \ac{cc} imaging~\cite{Sato2024ComptonNet}. Both 2D-DenseDeconNet and ComptonNet predict an image in a single forward pass without iterative updates. Reconstruction quality is measured using \ac{nmse} as a function of iteration/layer depth, while the \ac{cc} experiment additionally examines spatial cross-profiles. The unfolded architecture uses $K=10$ layers in both settings.

{\bf \ac{rl} experiment:}
We first consider synthetic photon-limited deconvolution. Ground-truth images are $32\times32$ pixels and contain randomly positioned and oriented soft-edged geometric shapes. Measurements are generated according to \eqref{eq:poisson_model} after zero-padded convolution with a normalized $9\times9$ Gaussian point-spread function, using a photon level of $P=5000$ and a known additive background. The dataset contains 3000 training and 1000 validation images, with 500 independently generated test images.
Classical and penalized \ac{mlem} are evaluated for up to 300 iterations. A representative reconstruction is shown in Fig.~\ref{fig:rl-thin-rectangle}. The ten-layer unfolded reconstruction is closer to the ground truth than direct 2D-DenseDeconNet and surpasses conventional iterative methods in both accuracy and runtime. This behavior is quantified in Fig.~\ref{fig:rl-convergence}: an \ac{nmse} of $0.11$ after ten unfolded layers, compared with $0.18$ for penalized \ac{mlem} after 300 iterations, and $0.27$ for DenseDeconNet, while the classical \ac{mlem} is non-converging. 

{\bf \ac{cc} experiment:}
We next consider \ac{cc} imaging using the GEANT4-based GAMOS framework~\cite{arce2008gamos,harkness2012compton}.
The detector consists of two $50\times50\times5$~mm CsI(Tl) planes separated by $50$~mm, with $7\%$ FWHM energy resolution. The source plane is reconstructed on an $81\times81$ grid with $5$~mm spacing, and the evaluated scene contains a rectangular $^{137}$Cs source.  The training, validation, and test sets contain $10{,}880$, $2{,}240$, and $500$ sets of 1000 events, respectively. The system matrix was constructed event by event ~\cite{Kim2007ComptonProjectors, Ren2025Attenuation} as $A_{ij}=\Delta_{ij}R_{ij}^{-2}\exp(-\mu_t L_{ij})$, where $\Delta_{ij}$ indicates whether any of the sampled rays on the measured Compton cone intersects pixel $j$, $R_{ij}$ is the distance from the first interaction point to the pixel center, $L_{ij}$ is the corresponding attenuation path length within the scatterer, and $\mu_t=0.035\,\mathrm{mm}^{-1}$. The iterative methods share the system matrix, while ComptonNet maps its measurement input directly to an image; classical and penalized \ac{mlem} are evaluated for up to 100 iterations.
Fig.~\ref{fig:cc-rectangle-reconstructions} shows representative reconstructions and their horizontal profiles at $y=0$. After ten updates, the conventional methods recover the approximate source location but produce broad responses and substantial intensity inside the rectangular structure. Additional iterations sharpen their reconstructions, although classical \ac{mlem} becomes increasingly nonuniform due to noise amplification of the model mismatch. The ten-layer unfolded reconstruction resolves the two target intersections more clearly than ComptonNet, whose filled-in rectangle produces only a shallow central dip in the $y=0$ profile.

The corresponding \ac{nmse} evolution in Fig.~\ref{fig:cc-nmse-convergence} further demonstrates the finite-depth advantage. The proposed method reaches an \ac{nmse} of approximately $0.15$ after ten layers, compared with roughly $0.35$-$0.4$ for the conventional methods after ten iterations. Penalized \ac{mlem} improves gradually to approximately $0.23$ after 100 iterations, whereas classical \ac{mlem} begins to deteriorate after approximately 15 iterations due to continued noise amplification. ComptonNet struggles with closely spaced geometries, achieving an NMSE of approximately $0.70$.
 The reduction in reconstruction depth also translates into lower measured inference time: as reported in
Table~\ref{tab:inference_complexity}.

\section{Conclusions}
% Restate the finite-depth Poisson inverse-problem formulation.
% Summarize the model-based/data-driven architecture.
% Summarize synthetic deconvolution and real-world Compton results.
% Emphasize only what is experimentally demonstrated. 
We proposed a deep-unfolded penalized \ac{mlem} framework for finite-depth Poisson image reconstruction, which preserves the analytical forward and backward projections and multiplicative update structure while learning iteration-dependent reconstruction parameters and a sensitivity correction. The resulting architecture combines model-based reconstruction with data-driven adaptation to improve recovery under limited iteration budgets and model mismatch. Experiments on photon-limited deconvolution and \ac{cc} imaging showed that the proposed method achieves improved reconstruction quality using substantially fewer update stages than iterative \ac{mlem} run and task-adapted direct models.

\bibliographystyle{IEEEtran}
\bibliography{IEEEabrv,refs}

\end{document}